\documentclass[12pt,a4paper]{article}
\usepackage[utf8]{inputenc}
\usepackage{amsmath}
\usepackage{amsfonts}
\usepackage{amssymb}
\usepackage{graphicx}
\usepackage{algorithm}
\usepackage{algpseudocode}
\usepackage{booktabs}
\usepackage{hyperref}
\hypersetup{
    colorlinks=true,
    citecolor=black,
    linkcolor=black,
    urlcolor=black
}
\usepackage{geometry}
\usepackage{setspace}
\usepackage{geometry}
\usepackage{xcolor}
\usepackage{natbib}
\usepackage{multirow}
\usepackage{array}

\title {Improving Linear Regression on Small Datasets via Gaussian Process and Extreme Value Theory Based Data Augmentation}

\author{M.I. Salay$^{1*}$ and S.G.J. Senarathne$^{2}$ \\
$^{1}$Department of Mathematics, University of Peradeniya \\
$^{2}$Department of Statistics and Computer Science, University of Peradeniya \\
$^{*}$isalay967@gmail.com}

\date{}

\begin{document}

\maketitle

\begin{abstract}
Small sample sizes pose significant challenges in regression analysis, often leading to violations of classical assumptions such as normality, homoscedasticity, and independence of residuals. These violations compromise parameter estimation accuracy, reduce statistical power, and limit the generalizability of findings. This study introduces the Gaussian Process-based Modified Extreme Value Theorem (GP-MEVT) method, a novel hybrid data augmentation approach that combines Gaussian Process with Extreme Value Theory to address these limitations. The GP-MEVT method generates augmented observations that extend the predictor space beyond the observed range while preserving the underlying linear structure and introducing controlled variability based on residual variation, through comprehensive simulation studies across three variance scenarios ($\sigma = 2, 5, 8$) and sample sizes ($n = 10, 15, 20$). Here, we demonstrate that GP-MEVT achieves a higher rate of assumption satisfaction, substantially outperforming standard bootstrap and bootstrap with noise methods. The proposed method also exhibits reasonable parameter estimation accuracy, with intercept and slope estimates consistently closer to true parameter values, and maintains competitive or superior model fitting performance as measured by root mean square error. Application to a real-world dataset confirms these advantages, with GP-MEVT achieving a 67.1\% assumption satisfaction rate compared to 17.3\% and 21.2\% for bootstrap alternatives. These findings establish GP-MEVT as a robust and reliable framework for fitting linear regression models to small datasets, offering practitioners a principled approach to statistical inference when sample size limitations are unavoidable.
\end{abstract}


\section{Introduction}

In various research domains, a recurrent aim is to achieve substantial sample sizes, particularly when employing intricate statistical models. However, challenges in data collection are common, especially in studies that involve limited population sizes or specific criteria. As noted by \citet{peto1976design}, larger sample sizes typically increase the likelihood of obtaining statistically significant results. Nonetheless, several obstacles, including research protocols, restricted populations, time limitations, and ethical considerations, can impede the recruitment of larger samples. As a result, researchers often face difficulties in making robust conclusions based on studies with limited sample sizes due to insufficient statistical power \citep{button2013power, price2012small, lee2004evaluation, scheines1999bayesian}. Despite these challenges, small datasets remain essential for drawing meaningful inferences, requiring the application of specialized methods to ensure accurate and reliable outcomes.

Simple Linear Regression (SLR) is a prominent statistical technique employed across diverse disciplines, which describes a linear relationship between an independent variable (predictor) and a dependent variable (response), thereby enabling researchers to formulate predictions and infer associations between variables. The establishment of a regression model is critical for elucidating the distribution of a dataset and identifying potential patterns within the data \citep{wooldridge2016introductory}. To ensure the validity of SLR results, certain assumptions about the underlying data must be fulfilled \citep{meuleman2015regression}. Firstly, SLR presumes a linear relationship between the dependent and independent variables, meaning that predictions for $y$ remain consistent across all ranges of $x$. Secondly, it is assumed that the residuals are normally distributed, with a mean of zero and constant variance (homoscedasticity). Additionally, the residuals should be independent of one another. In cases where multiple predictors are employed, which is known as the Multiple Linear Regression, it is equally crucial to assume that these predictor variables are uncorrelated (multicollinearity) \citep{fox2015applied, montgomery2012dirt, johnson2011probability}. Adhering to these assumptions is crucial for ensuring the accuracy and interpretability of model outcomes, as deviations from them may markedly impair model performance \citep{harrell2001regression}. However, in numerous applications, these assumptions are frequently violated \citep{woolson2002statistical}, leading to potential parameter bias and overestimation of variance components, which can compromise research findings and lead to erroneous conclusions. The underlying assumptions of statistical models are essential for hypothesis testing and interval estimation. It is crucial to approach the validity of these assumptions with a degree of skepticism and to conduct thorough analyses to assess the model's adequacy. The inadequacies discussed can have significant implications for the reliability of the model. Substantial violations of these assumptions may lead to an unstable model, wherein different samples could result in divergent models yielding contradictory conclusions. Moreover, reliance on standard summary statistics such as $t$ or $F$ statistics and R-squared ($R^2$) often fails to detect deviations from the fundamental assumptions. These statistics represent ``global'' properties of the model, and, as such, they do not provide assurance of model adequacy.

The significance of sample size is crucial in the context of fitting regression models. Larger sample sizes typically yield more dependable estimates of model parameters, whereas smaller sample sizes can produce biased estimates, lower statistical power, and challenges in interpreting the variance of the dependent variable \citep{tabachnick2013using, cohen2013statistical}. In particular, when working with limited datasets, the accuracy of slope and intercept estimates may be compromised, thereby increasing the risk of overfitting and constraining the generalizability of the results \citep{button2013power}. These challenges emphasize the necessity for specialized methodologies to enhance regression modeling under such conditions. When dealing with small datasets, certain assumptions are more likely to be violated. This can lead to problems such as non-normal residuals and heteroscedasticity \citep{mcdonald2014handbook}. Additionally, overfitting where the model learns patterns too closely from a small number of data points can significantly affect model performance, causing it to capture noise rather than the true underlying relationships, as highlighted by \citet{hastie2009elements}. This issue is especially pronounced with limited data, resulting in a high variance in model estimates and consequently unreliable predictions \citep{james2013introduction}. \citet{dietterich1995overfitting} further underscored the complexity generalization trade-off, indicating that maintaining a balance between model accuracy and overfitting becomes increasingly challenging when working with small datasets, where the model may either overfit or struggle to generalize effectively.

To address the challenges associated with small datasets, researchers utilize several advanced techniques. Bayesian analyses stand out as they do not depend on the assumption of large sample sizes, unlike maximum likelihood (ML) estimation. It incorporates prior knowledge to reduce overfitting, thereby providing a more robust foundation for inference \citep{gelman2013bayesian}. This characteristic facilitates the analysis of typically smaller datasets without a significant loss of statistical power while maintaining precision. As illustrated by \citet{lee2004evaluation}, Bayesian estimation achieves this with a substantially smaller ratio of parameters to observations. In their study, a ratio of 1:3 is adequate, in contrast to the 1:5 ratio that is often required for ML estimation. Furthermore, \citet{hox2012few} conducted a simulation study that demonstrated how Bayesian estimation allows for the effective use of smaller datasets compared to ML estimation.  However, appropriate prior distributions must be applied to the parameters; otherwise, \citet{mcneish2016using} noted that Bayesian estimates may end up being less accurate and less efficient than those derived from a frequentist approach.

The most popular method for handling small samples is by generating new data points from the existing sample. Among data augmentation techniques, Bootstrap resampling is the easiest and most popular method among statisticians. Bootstrap resampling generates multiple augmented datasets, allowing researchers to estimate the stability and variability of model predictions \citep{efron1994introduction}. Bootstrap and Jackknife are resampling techniques that make no assumptions when estimating the parameter $\beta$. In their research, \citet{sahinler2007bootstrap} compared these two methods and discussed strategies for constructing a regression model using both. They concluded that a larger number of bootstrap replicates leads to parameter estimates that are closer to the true values. They also recommended using 1000 bootstrap replicates to effectively estimate the variance and standard errors (SE). Their study assessed the accuracy of the Bootstrap and Jackknife methods for estimating the distribution of regression parameters across various sample sizes and bootstrap replicates. They found that the bootstrap method is suitable for LR, even when the error distribution is not normal. Another notable advantage of bootstrap approximations is their ability to provide parameter estimates with potentially smaller sample sizes compared to ordinary least squares. However, several disadvantages of bootstrap methods have been highlighted in the literature by \citet{ma2018probabilistic, wan2013hybrid, nelson2014cluster, phaladiganon2011bootstrap}. Key drawbacks include the potential for the bootstrap distribution to inaccurately approximate the underlying distribution when the sample size is small and outliers are present. Furthermore, bootstrap methods are generally not recommended for data exhibiting dependency structures, such as time series data. Finally, the residual bootstrap approach is often not preferable when the fundamental assumptions of the underlying model are violated.

In recent times, there has been a growing interest in applying machine learning (ML) to address challenges associated with small datasets. A critical factor contributing to the success in this area is the ability to accurately estimate behavior in unknown domains by quantitatively learning patterns from a sufficient number of training examples. However, in cases involving smaller datasets that undeniably impact the development of ML models, researchers have found that the model fitting accuracy of ML models improved systematically with an increase in the training set size. Similarly, \citet{schmidt2017predicting} reported that the prediction error for the formation energy of perovskite compounds decreased monotonically as the size of the training set increased, following a power law; notably, doubling the training set resulted in a reduction of approximately 20\% in the error rate. \citet{lee2016prediction} explored ML models for the band gaps of inorganic compounds and discovered that the model fitting accuracy of ordinary least squares regression and LASSO models converged at certain training set sizes. Collectively, these studies clearly illustrate that the limited availability of training data complicates the detection of patterns. Neural networks (NN), when integrated with suitable regularization techniques, are capable of effectively discerning complex relationships even within small datasets \citep{goodfellow2016deep}. These techniques, in conjunction with cross-validation and ensemble models, improve generalization and robustness, thereby addressing the challenges posed by limited data. However, as is the case in ML, developing a model becomes more challenging when the training dataset is small. Additionally, the computational resources required for executing ML and NN can be substantial, contributing to increased costs of the study. 

Gaussian Process (GP) offers a strong alternative by adjusting its complexity in response to the observed data. This built-in adaptability allows GPs to effectively capture complex patterns and relationships, even with limited observations. As a result, they can effectively tackle the challenges associated with small sample sizes in regression model fitting. As opposed to compelling the data to fit a rigid, pre-defined functional form, GPs establish a probability distribution over a space of possible functions, rendering them particularly sensitive to the subtleties embedded within the data \citep{goldberg1997regression}. The GP structure imposed by the selected covariance function serves as a rigorous form of regularization, mitigating the risk of overfitting the limited training data. The kernel function effectively encodes assumptions related to the smoothness and overall shape of the underlying function, thereby directing the model's learning trajectory in the face of minimal data points. Building upon the foundational principles of GPs, Gaussian Process Regression (GPR) emerges as a tool for data-driven modeling. In contrast to traditional numerical approaches and NNs, GPR transcends the conventional paradigms predicated on model parameter selection; instead, it derives the posterior probability distribution in accordance with Bayesian principles, leveraging the premise that the prior distribution of observations conforms to a normal distribution \citep{rasmussen2003gaussian, schulz2018tutorial}. Moreover, GPR necessitates significantly fewer samples than neural networks, thereby alleviating computational burdens. Nevertheless, it is important to note that GPR may exhibit performance degradation when faced with inappropriate volumes of regression data or overly complex relationships \citep{wang2016knn}.

The objective of this research is to propose a method for fitting a regression model to small samples in a way that more reliably adheres to the classical regression assumptions. Recognizing both the capabilities and limitations of GPR, particularly its sensitivity to data volume and complexity, this study develops an enhanced methodology that leverages GP's strengths while mitigating its weaknesses through strategic data augmentation. This motivated approach addresses the critical challenge of working with small sample sizes where traditional regression methods often fail to satisfy fundamental statistical assumptions.

\section{Proposed Regression Framework for Small Sample Analysis}

The study proposes a hybrid data augmentation technique called the Gaussian Process-based Modified Extreme Value Theorem (GP--MEVT) method. This approach integrates Gaussian Processes (GPs) with Extreme Value Theory (EVT) to generate realistic augmented samples that extend the observed data range while preserving local structure. The primary focus of this approach is to address the issue of heteroscedasticity and the non-normality in residuals of smaller datasets. By combining the augmented and original datasets, this method specifically targets correcting heteroscedasticity and non-normality, thereby enhancing the original sample and increasing the likelihood that the fitted regression model will meet classical regression assumptions.

The GP-MEVT methodology addresses a critical challenge in data-driven modeling where traditional approaches struggle with limited sample sizes. Insufficient training data typically yields imprecise system characterization and creates two fundamental problems: uncertainty regarding the true population domain and ambiguity in how data gaps should be filled. The GP-MEVT framework directly confronts these limitations by strategically combining GP's ability to model uncertainty and make predictions with minimal data, with EVT capability to extrapolate beyond observed ranges. This hybrid approach systematically estimates the input data domains and generates augmented samples to fill data gaps while simultaneously ensuring that the augmented dataset satisfies classical regression assumptions. GP-MEVT not only augments the sample size but also specifically corrects for violations of homoscedasticity and normality, thereby producing statistically robust models from limited observations, that is a crucial advantage when data acquisition is constrained by practical or economic factors.

\subsection{Gaussian Process Modeling}

Let $(x_i, y_i)_{i=1}^{n}$ represent the observed data, where $y_i$ denotes the response variable and $x_i$ the corresponding predictor. A GP model was first fitted to characterize the functional relationship between $x$ and $y$. The model assumes that the response follows:
\begin{equation}
y(x) = f(x) + \varepsilon
\end{equation}
where $f(x) \sim \text{GP}(0, k(x, x'))$ denotes a GP prior with kernel function $k(x, x')$, and $\varepsilon \sim \mathcal{N}(0, \sigma^2)$ represents Gaussian noise. The kernel defines the smoothness and correlation between observations. In this analysis, the linear kernel was used to adopt the linear relationship between predictor and response variables. After fitting the GP model, residuals were computed as $e_i = y_i - \hat{y}_i$, and the residual standard deviation (RSD) was obtained using:
\begin{equation}
\text{RSD} = \sqrt{\frac{1}{n}\sum_{i=1}^{n}(y_i - \hat{y}_i)^2}
\end{equation}
This measure quantifies the level of random noise present in the data, which is subsequently used to introduce variability during augmentation.

\subsection{Determination of Extreme Boundaries via MEVT}

To define an extended domain for generating augmented predictors, boundary limits were estimated using the Modified Extreme Value Theorem (MEVT) method which was introduced by \citet{inproceedings}. The predictor variable $x$ was divided at its midpoint into left ($L$) and right ($R$) subsets. For each subset, the sample mean ($\bar{L}$, $\bar{R}$) and standard deviation ($s_L$, $s_R$) were computed. Using the constants ($k_1$ and $k_2$) proposed by \cite{lowery1970comparison} and an exceedance probability $H_p$, the lower and upper boundary limits ($a$, $b$) were obtained as:
\begin{equation}
\begin{split}
a &= (k_1 s_L) + (\bar{L} - k_2 s_L) \ln[-\ln(H_p)] \\
b &= (k_1 s_R) + (\bar{R} - k_2 s_R) \ln[-\ln(H_p)]
\end{split}
\end{equation}
These boundaries enable the model to extrapolate beyond the observed range without producing implausible predictor values.

\subsection{Data Augmentation via GP Sampling}

A total of $n_{aug}$ new predictor values $x_{aug}$ were uniformly sampled from the interval $[a, b]$, $\{x_{aug_j}\}_{j=1}^{N_{aug}} \sim \text{Uniform}(a, b)$. For each $x_{aug}$, the GP model predicted the corresponding mean response $\hat{y}_{GP}(x_{aug})$. To mimic natural variation, Gaussian noise scaled by the residual variability was added:
\begin{equation}
y_{aug} = \hat{y}_{GP}(x_{aug}) + \mathcal{N}(0, (\text{RSD}))
\label{eq:y_aug}
\end{equation}
The generated $(x_{aug}, y_{aug})$ pairs were then combined with the original dataset to create an augmented sample $D_{combined} = D_{sample} \cup \{(x_{aug_j}, y_{aug_j})\}_{j=1}^{N_{aug}}$.

\subsection{Model Re-Estimation and Diagnostic Evaluation}

Finally a SLR model was refitted using the combined dataset to assess whether data augmentation improved parameter stability and model assumption satisfaction. Diagnostic tests of normality, homoscedasticity, independence, and linearity were conducted using a specific function to evaluate the reliability of the augmented model.

For all computational results, the constants $a$ and $b$ were derived by fitting the parameters as $k_1 = {\frac{\sqrt6}{\pi}}$ and $k_2 = 0.450041$. The exceedance probability was set at $H_p = 0.025$.

\subsection{GP-MEVT Algorithm}
The GP-MEVT algorithm addresses small sample regression problems by combining GP modeling with EVT for intelligent data augmentation. It first attempts standard regression, and if assumptions fail, fits a GP model to the limited data, then uses the MEVT to identify plausible boundary regions in the predictor space. The algorithm generates data points within these statistically justified boundaries by sampling from the GP's predictive distribution with added noise scaled to match residual variation, then re-fits the regression model on the augmented dataset. This approach strategically expands the sample in regions where additional data is most likely to improve model reliability and satisfy regression assumptions.

\begin{algorithm}
\caption{: GP--MEVT}
\begin{algorithmic}[1]
\State \textbf{Input:} Small sample dataset: $D_{sample} = \{(x_i, y_i)\}_{i=1}^{n}$
\State \textbf{Input:} Assumption verification procedure: $AVP$
\State \textbf{Input:} Number of augmented data points: $N_{aug}$
\State \textbf{Input:} Exceedance probability: $H_p = 0.025$
\State \textbf{Input:} $k_1 = \sqrt{6}/\pi$, $k_2 = 0.450041$
\State \textbf{Step 1: Fit initial regression model}
\State Fit a traditional regression model $M_{reg}$ on sample $D_{sample}$.
\State \textbf{Step 2: Check regression assumptions}
\State Apply the assumption verification procedure $AVP$ on $M_{reg}$.
\If{assumptions are satisfied}
    \State \textbf{Output:} $M_{reg}$ (STOP)
\Else
    \State Proceed to Step 3.
\EndIf
\State \textbf{Step 3: Gaussian Process modeling}
\State Fit a Gaussian Process model $M_{GP}$ on $D_{sample}$ with kernel function $k(x, x') = x^T x'$.
\State \textbf{Step 4: Estimate residual variation}
\State Compute residuals: $e_i = y_i - \hat{y}_i$ for $i = 1, \ldots, n$
\State Calculate residual standard deviation (RSD).
\State \textbf{Step 5: Determine extreme boundaries via MEVT}
\State Divide predictor variable $X$ at its midpoint into left ($L$) and right ($R$) subsets.
\State Compute sample mean and standard deviation for each subset: $(\bar{L}, s_L)$ and $(\bar{R}, s_R)$
\State Calculate lower boundary:
\State $a = (k_1 s_L) + (\bar{L} - k_2 s_L) \ln[-\ln(H_p)]$
\State Calculate upper boundary:
\State $b = (k_1 s_R) + (\bar{R} - k_2 s_R) \ln[-\ln(H_p)]$
\State \textbf{Step 6: Data augmentation via GP sampling}
\State Generate $N_{aug}$ predictor values uniformly from interval $[a, b]$.
\For{$j = 1$ to $N_{aug}$}
    \State Predict mean response using GP model: $\hat{y}_{GP}(x_{aug_j})$
    \State Add Gaussian noise scaled by RSD.
\EndFor
\State Combine original and augmented data.
\State \textbf{Step 7: Model re-estimation and diagnostic evaluation}
\State Fit a new regression model $M_{reg}^{combined}$ on $D_{combined}$.
\State Apply assumption verification procedure $AVP$ on $M_{reg}^{combined}$.
\If{all assumptions are satisfied}
    \State \textbf{Output:} Final model $= M_{reg}^{combined}$ 
\Else
    \State \textbf{Output:} Model requires further refinement
\EndIf
\end{algorithmic}
\end{algorithm}

A comprehensive assumption-checking procedure $AVP$ was established to evaluate the three critical assumptions of liner regression. This procedure serves as the foundation for validating the regression model. The assumptions of homoscedasticity, normality of residuals, and independence of residuals are evaluated through specific statistical tests, accepting the assumption that at least one of these tests indicates independence with p-values exceeding 0.05. A model is deemed valid only when all three assumptions are satisfied simultaneously. This multi-test strategy ensures a robust assessment by avoiding reliance on a single statistical test, which may be sensitive to sample size or specific data characteristics. For each run of the simulation studies, it is recorded whether all assumptions are satisfied or if any assumptions are violated.

\section{Example 1: Simulation Study}

A simulation framework was employed to further validate the proposed method. For the simulation study, a predictor variable $X$ was randomly sampled from a uniform distribution ranging from -1 to 1. The linear relationship between $X$ and the response variable $Y$ is characterized by an intercept of $\beta_0 = 5$ and a slope of $\beta_1 = 10$. Accordingly, the response variable $Y$ is generated from a normal distribution, $N(E(Y),\sigma^2)$, where  $E(Y)=\beta_0+\beta_1 X$ and $\sigma$ is the residual standard deviation (as in Equation \ref{eq:y_aug}). To obtain robust results and enhance the generalizability of the findings, three scenarios were considered by varying the value of $\sigma$, which determines the strength of the linear relationship between $X$ and $Y$ (see Table \ref{Tab1}). 

\begin{table}[H]
\caption{Three Simulation Scenarios}
\begin{center}
\begin{tabular}{ccc}
\hline
Scenario & Residual Standard Deviation ($\sigma$) & $R^2$ \\ \hline
1        & 8                                                   &  0.34         \\
2        & 5                                                   &  0.57         \\
3        & 2                                                   &  0.90         \\ \hline
\end{tabular}
\label{Tab1}
\end{center}
\end{table}

For each scenario, samples of sizes 10, 15, and 20 were considered. A total of 1,000 simulated datasets were generated for each scenario and sample size, selecting only the datasets in which at least one regression assumption failed. Additionally, an augmented dataset of sizes 10, 15, and 10 were generated, respectively, for the derived samples. This ensured that the augmented sample sizes did not exceed the original sample size and that the total sample size, comprising both the initial and augmented samples, was close to 30.

In the simulation study, the proposed method was evaluated against the bootstrap method and the bootstrap method with added noise to each resample. The noise level for the Bootstrap method was fixed at 0.5 for all cases. Notably, the bootstrap with noise approach generated unique augmented data, whereas the standard bootstrap method repeatedly populated the sample with the same values. An LR model was fitted for each method, considering different sample sizes and varying population standard deviations. Primarily, the acceptance probabilities of the models were compared while monitoring the $R^2$ values, the estimated parameters \& their SEs, and the root mean square error (RMSE) of the fitted models.

The simulation findings indicated that population standard deviation exerted a substantive influence on model performance. As this parameter decreased, both model fit indices and parameter precision improved considerably, as evidenced by the progressive reduction of standard errors for intercept and slope estimates. Furthermore, slope coefficients consistently approximated 10, and intercept coefficients remained in close proximity to 5 across all scenarios and augmentation strategies.

\begin{figure}[H]
\centering
\vspace{1cm}
\begin{tabular}{cc}
\includegraphics[width=0.45\textwidth]{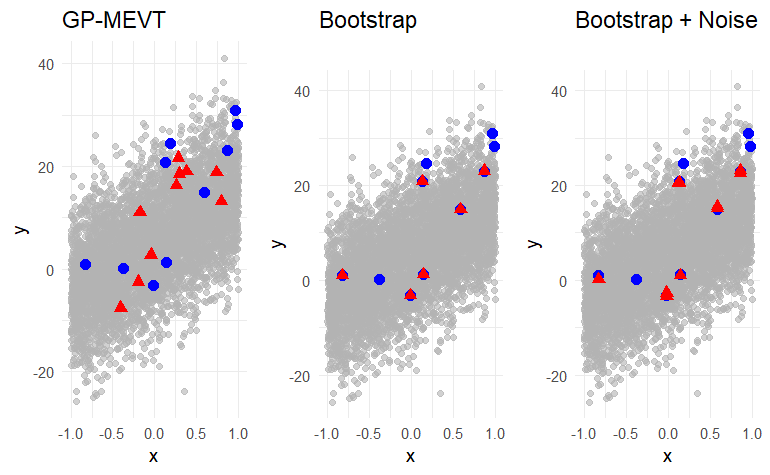} &
\includegraphics[width=0.45\textwidth]{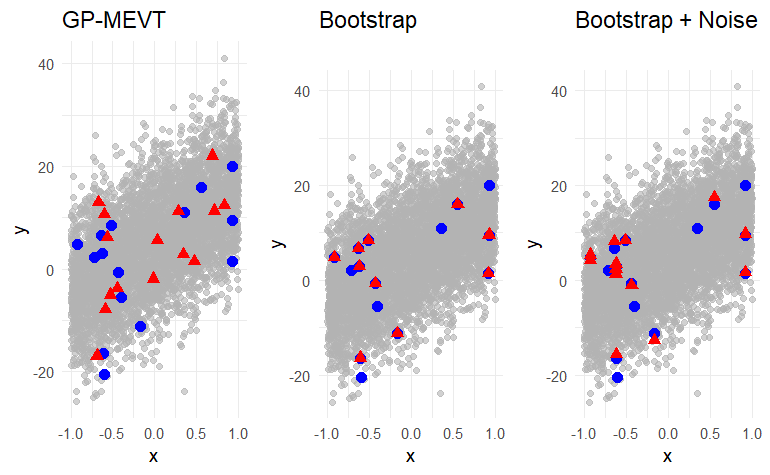} \\
(a) Sample of size 10 & (b) Sample of size 15 \\[0.1cm]
\end{tabular}
\vspace{0.5cm}
\begin{tabular}{cc}
\includegraphics[width=0.45\textwidth]{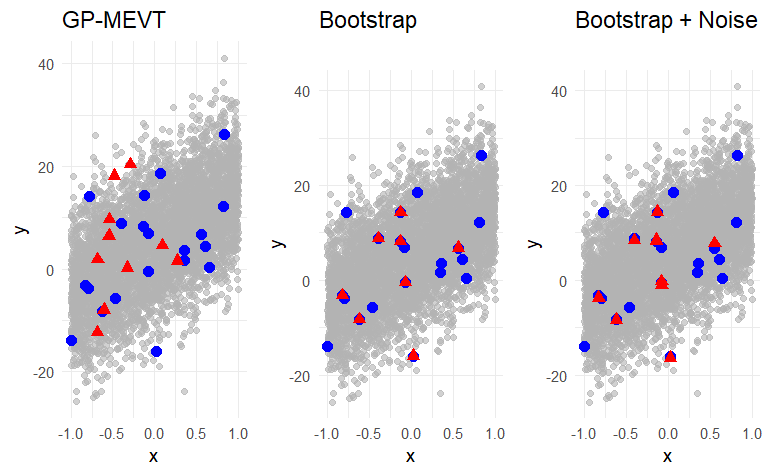} &
\includegraphics[width=0.30\textwidth]{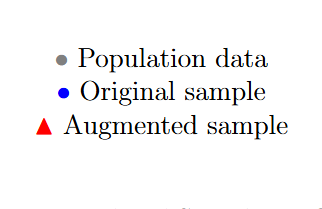} \\
(c) Sample of size 20 & 
\end{tabular}
\caption{Augmented Data Distribution for a Random Sample in GP-MEVT, Bootstrap, and Bootstrap with Noise methods for Scenario 1.}
\label{Fig1}
\end{figure}

\begin{figure}[H]
\centering
\vspace{0.1cm}
\begin{tabular}{cc}
\includegraphics[width=0.42\textwidth]{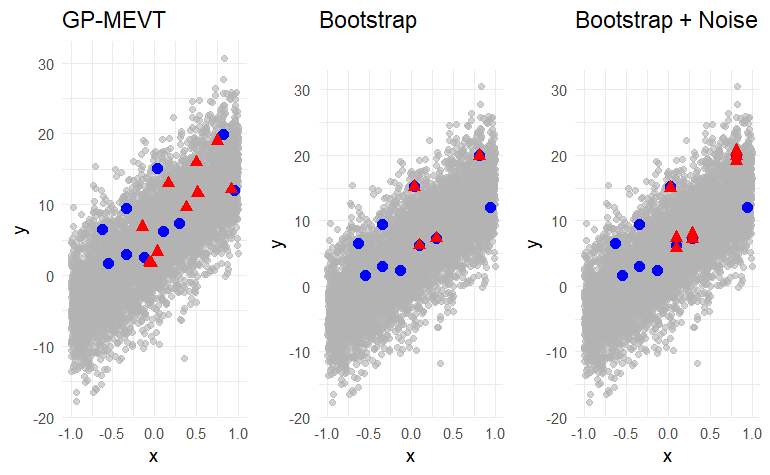} &
\includegraphics[width=0.42\textwidth]{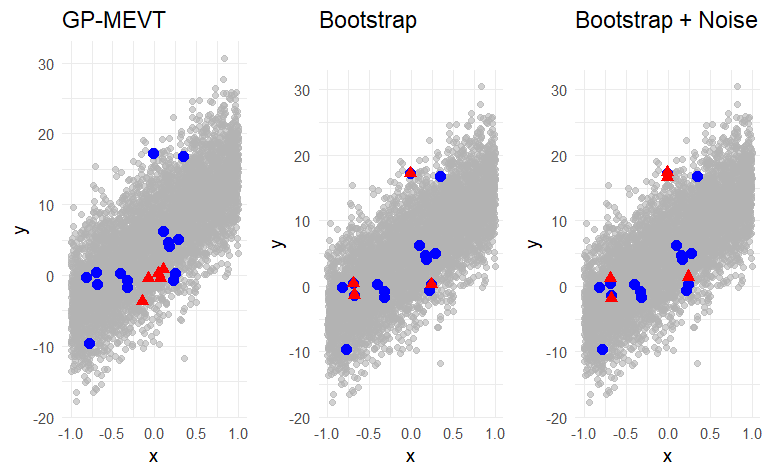} \\
(a) Sample of size 10 & (b) Sample of size 15 \\[0.1cm]
\end{tabular}
\vspace{0.5cm}
\begin{tabular}{cc}
\includegraphics[width=0.42\textwidth]{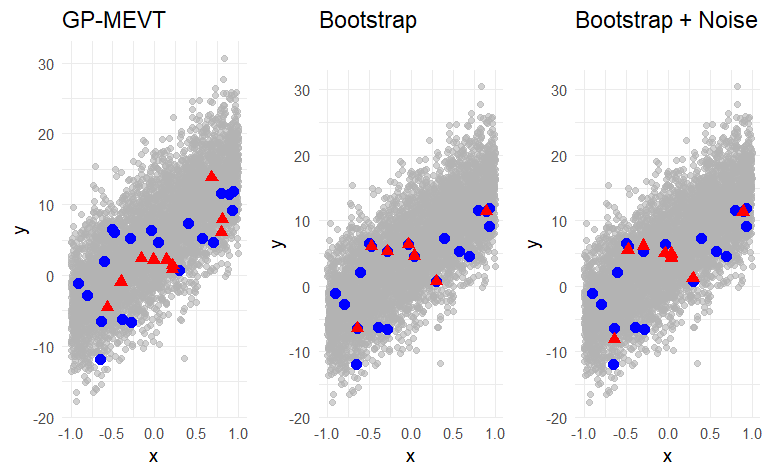} &
\includegraphics[width=0.30\textwidth]{Legend.png} \\
(c) Sample of size 20 & 
\end{tabular}
\caption{Augmented Data Distribution for a Random Sample in GP-MEVT, Bootstrap, and Bootstrap with Noise methods for Scenario 2.}
\label{Fig2}
\end{figure}

\begin{figure}[H]
\centering
\vspace{0.5cm}
\begin{tabular}{cc}
\includegraphics[width=0.42\textwidth]{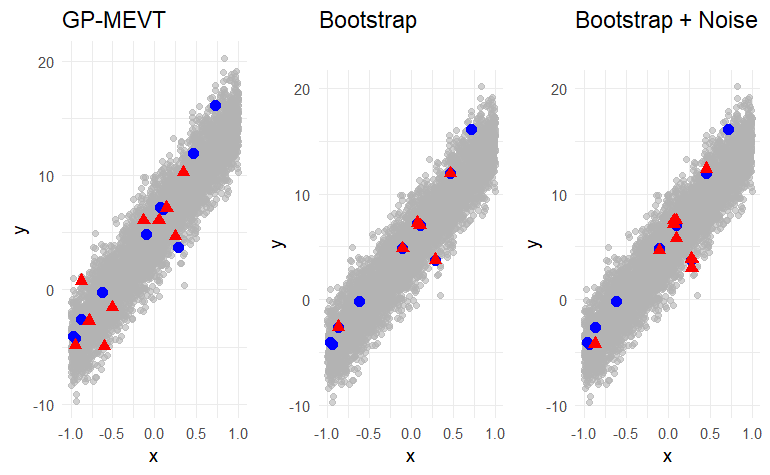} &
\includegraphics[width=0.42\textwidth]{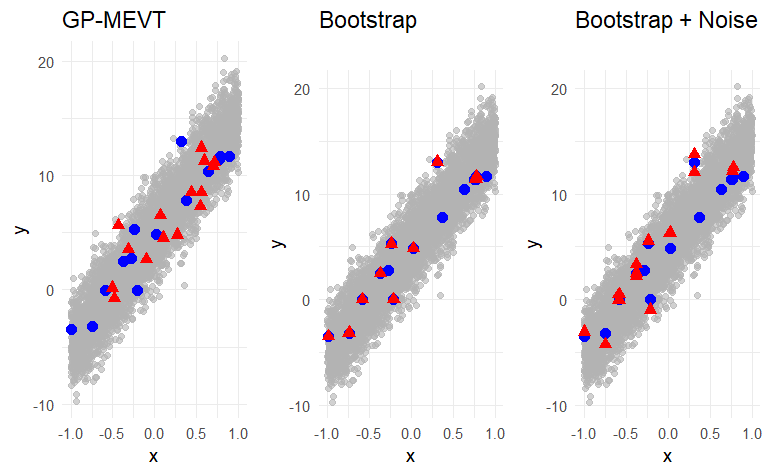} \\
(a) Sample of size 10 & (b) Sample of size 15 \\[0.1cm]
\end{tabular}
\vspace{0.5cm}
\begin{tabular}{cl}
\includegraphics[width=0.42\textwidth]{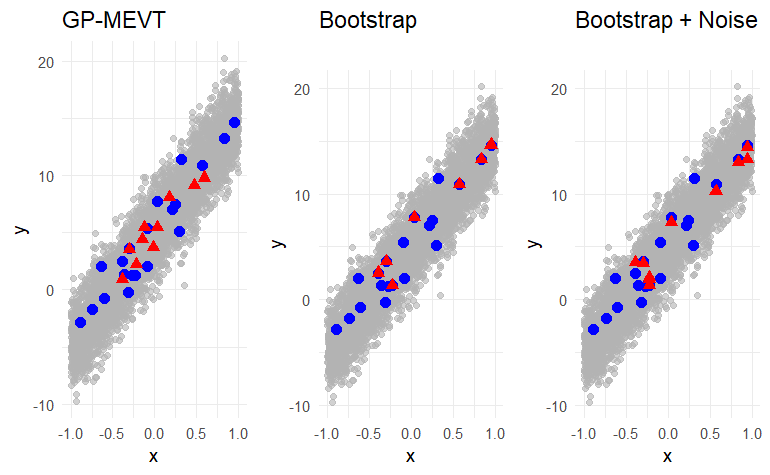} &
\includegraphics[width=0.30\textwidth]{Legend.png} \\
(c) Sample of size 20 & 
\end{tabular}
\caption{Augmented Data Distribution for a Random Sample in GP-MEVT, Bootstrap and Bootstrap with Noise methods for Scenario 3.}
\label{Fig3}
\end{figure}

Figures \ref{Fig1}, \ref{Fig2}, and \ref{Fig3} show how each method generates augmented data to improve model fit during a single simulation run. These figures illustrate that Bootstrap methods increase the number of data points simply by duplicating them or selecting data points close to the sample data, while GP-MEVT extends beyond the sample range to generate data within the population range, where the sample does not adequately represent the population.  The corresponding summary statistics for each scenario are provided in \ref{tab:case1_single}, \ref{tab:case2_single}, \ref{tab:case3_single} in the Appendix.

The computational results for all 1000 simulations for each scenario are summarized in Table \ref{tab:all_cases_combined}. As can be seen in Table  \ref{tab:all_cases_combined}, the most notable advantage of GP-MEVT lies in its ability to produce models that satisfy regression assumptions far more consistently than the competing methods. Across all three variance scenarios and every sample size tested, GP-MEVT consistently achieved assumption-passing rates that were three to four times higher than those of standard Bootstrap and Bootstrap with Noise approaches. This pattern held remarkably stable regardless of whether the underlying data exhibited high, moderate, or low variance, demonstrating that GP-MEVT's superiority in maintaining statistical validity is not dependent on the specific characteristics of the dataset. While both Bootstrap methods struggled to generate augmented samples that met the necessary statistical conditions for reliable inference, GP-MEVT's boundary-based sampling strategy proved substantially more effective at preserving the fundamental assumptions required for valid regression modeling. The consistently low passing rates of bootstrap-based methods suggest that simply resampling or adding random noise does not adequately address the fundamental assumption violations present in the original small samples. GP-MEVT's success in this regard indicates that its data augmentation approach effectively reconstructs data that satisfy specially the normality and homoscedasticity requirements of classical LR. Notably, assumption-passing rates tend to decrease with increasing sample size for GP-MEVT, while bootstrap methods show more variable patterns. This may indicate that larger initial samples place tighter restrictions on the augmentation process, which makes it harder to fix underlying distributional problems (Assumption Satisfaction).

In terms of parameter estimation accuracy, all three methods demonstrated reasonable performance in recovering the true regression coefficients, with estimates consistently falling within a narrow range around both the true values and the estimates obtained from the full dataset. However, GP-MEVT showed an advantage in producing parameter estimates that are more closely approximated to the parameter estimates of the population data across all variance scenarios. This pattern was evident for both intercept and slope coefficients.

The comparison of model fit statistics reveals an interesting pattern in how the different methods capture the true underlying relationship between variables. While all three approaches produced relatively similar $R^2$ values overall, the Bootstrap methods consistently exhibited a tendency toward slightly higher explanatory power, particularly in high-variance scenarios. However, this apparent advantage may actually reflect a weakness rather than a strength, as GP-MEVT's $R^2$ values aligned more closely with the population benchmark from the full dataset, suggesting that the Bootstrap methods were potentially overfitting to sample-specific noise rather than capturing the genuine population relationship. This distinction became lesser in moderate and low variance scenarios, where the population signal was stronger and all methods converged toward similar $R^2$ values that more accurately reflected the true explanatory power. The pattern suggests that GP-MEVT helps prevent the artificial inflation of model fit that can occur when augmented data too closely mirrors sample-specific peculiarities rather than population-level patterns. 

The mean RMSE values offer a more direct measure of model fitting accuracy. GP-MEVT consistently demonstrates competitive or superior RMSE across all scenarios. The lower RMSE values of GP-MEVT, combined with its superior assumption-passing rates and parameter accuracy, indicate that the augmented data generated by this method supports models that not only meet statistical validity requirements but also generalize effectively to unseen observations as well. Examining performance across the three variance scenarios reveals important insights. As population variance increases from scenario 3 to scenario 1, the relative advantage of GP-MEVT becomes more pronounced, particularly in assumption satisfaction. This suggests that GP-MEVT is especially valuable in high-noise environments where traditional regression assumptions are more easily violated and where small samples are particularly problematic.

\begin{table}[H]
\centering
\caption{Results for All 1000 Simulation Runs}
\vspace{0.5cm}
\footnotesize
\begin{tabular}{llccccccc}
\toprule
Sample & Data & Senario & Assumption & Mean & Mean & Intercept & Slope \\
Size & augmentation &  & satisfaction & R2 & RMSE & (Std. Err.) & (Std. Err.) \\
& method & & rate & & & & \\
\midrule
\multirow{9}{*}{10} & \multirow{3}{*}{GP-MEVT} & 1& 0.657 & 0.343 & 8.899 & 5.282 (1.707) & 9.970 (3.492) \\
& & 2& 0.644 & 0.536 & 5.534 & 5.178 (1.070) & 9.984 (2.193) \\
& & 3& 0.646 & 0.868 & 2.214 & 5.071 (0.428) & 9.983 (0.877) \\
\cmidrule{2-8}
& \multirow{3}{*}{Bootstrap} & 1& 0.169 & 0.389 & 8.904 & 5.268 (1.685) & 9.989 (3.067) \\
& & 2& 0.162 & 0.592 & 5.552 & 5.164 (1.056) & 10.033 (1.924) \\
& & 3& 0.160 & 0.895 & 2.221 & 5.066 (0.423) & 10.013 (0.770) \\
\cmidrule{2-8}
& \multirow{3}{*}{Bootstrap + Noise} & 1& 0.199 & 0.394 & 8.917 & 5.229 (1.689) & 10.163 (3.088) \\
& & 2& 0.208 & 0.593 & 5.540 & 5.135 (1.051) & 10.027 (1.922) \\
& & 3& 0.282 & 0.892 & 2.218 & 5.056 (0.428) & 10.013 (0.783) \\
\midrule
\multirow{9}{*}{15} & \multirow{3}{*}{GP-MEVT} & 1& 0.575 & 0.306 & 8.532 & 5.207 (1.404) & 9.741 (2.901) \\
& & 2& 0.676 & 0.530 & 5.333 & 5.138 (0.858) & 9.973 (1.770) \\
& & 3& 0.578 & 0.871 & 2.133 & 5.055 (0.343) & 9.983 (0.708) \\
\cmidrule{2-8}
& \multirow{3}{*}{Bootstrap} & 1& 0.142 & 0.361 & 8.555 & 5.167 (1.395) & 9.712 (2.516) \\
& & 2& 0.119 & 0.593 & 5.341 & 5.155 (0.861) & 10.006 (1.538) \\
& & 3& 0.119 & 0.898 & 2.136 & 5.062 (0.344) & 10.002 (0.615) \\
\cmidrule{2-8}
& \multirow{3}{*}{Bootstrap + Noise} & 1& 0.163 & 0.360 & 8.556 & 5.255 (1.402) & 9.728 (2.521) \\
& & 2& 0.157 & 0.591 & 5.338 & 5.161 (0.867) & 10.030 (1.548) \\
& & 3& 0.217 & 0.894 & 2.137 & 5.065 (0.352) & 10.009 (0.629) \\
\midrule
\multirow{9}{*}{20} & \multirow{3}{*}{GP-MEVT} & 1& 0.471 & 0.321 & 8.345 & 5.069 (1.433) & 9.913 (2.791) \\
& & 2& 0.471 & 0.535 & 5.216 & 5.043 (0.895) & 9.944 (1.744) \\
& & 3& 0.471 & 0.876 & 2.086 & 5.017 (0.358) & 9.975 (0.698) \\
\cmidrule{2-8}
& \multirow{3}{*}{Bootstrap} & 1& 0.181 & 0.358 & 8.356 & 5.059 (1.428) & 9.927 (2.542) \\
& & 2& 0.181 & 0.578 & 5.222 & 5.037 (0.893) & 9.955 (1.589) \\
& & 3& 0.181 & 0.894 & 2.089 & 5.015 (0.357) & 9.982 (0.636) \\
\cmidrule{2-8}
& \multirow{3}{*}{Bootstrap + Noise} & 1& 0.212 & 0.356 & 8.353 & 5.145 (1.436) & 9.880 (2.552) \\
& & 2& 0.221 & 0.574 & 5.221 & 5.090 (0.899) & 9.924 (1.597) \\
& & 3& 0.273 & 0.891 & 2.089 & 5.036 (0.363) & 9.968 (0.646) \\
\bottomrule
\end{tabular}
\label{tab:all_cases_combined}
\end{table}

\section{Example 2: Application}

To further validate the practical applicability of the GP-MEVT method beyond controlled simulation environments, we obtained a real-world dataset from the online repository (Kaggle), comprising 1,000 observations with a documented linear relationship between predictor and response variables ($R^2 = 0.8036$). This dataset was used to implement the typical scenario in applied research where a substantial population exists but researchers may only have access to a small subset due to practical constraints. To simulate such a resource limited situation, we deliberately selected a random sample of size $n = 10$ from this population, mimicking instances where investigators must conduct regression analysis with limited data, a common challenge in fields such as medical research with rare diseases, pilot studies in experimental sciences, or preliminary analyses in exploratory research. The estimated parameters for the whole dataset are $\beta_0 = 7.8739$ (SE = 0.0354) and $\beta_1 = 3.4226$ (SE = 0.0535), providing a known benchmark against which to evaluate the performance of different augmentation methods. First, as in example 1, the augmented data distribution for a random single sample of size 10 was evaluated. Table \ref{tab:secondary_sample} summarizes the regression summary statistics for the random sample, which was obtained from the secondary dataset.

\begin{table}[h!]
\centering
\caption{Regression Summary Statistics for the Random Sample}
\vspace{0.5cm}
\begin{tabular}{lr}
\toprule
Sample Data & \\
\midrule
Size & 10 \\
R-squared & 0.8753 \\
All assumptions satisfied & FALSE \\
Estimated intercept (SE) & 6.3537(2.5381) \\
Estimated slope (SE) & 3.9873(0.4978) \\
\bottomrule
\end{tabular}
\label{tab:secondary_sample}
\end{table}

Figure 4 illustrates the distribution of data-augmentation methods for a random sample of size 10 as detailed in Table \ref{tab:secondary_sample}. Here, the obtained sample violates regression assumptions, unlike the full population.

\vspace{0.5cm}
\begin{figure}[H]
\begin{center}
\begin{tabular}{cc}
\includegraphics[width=10cm]{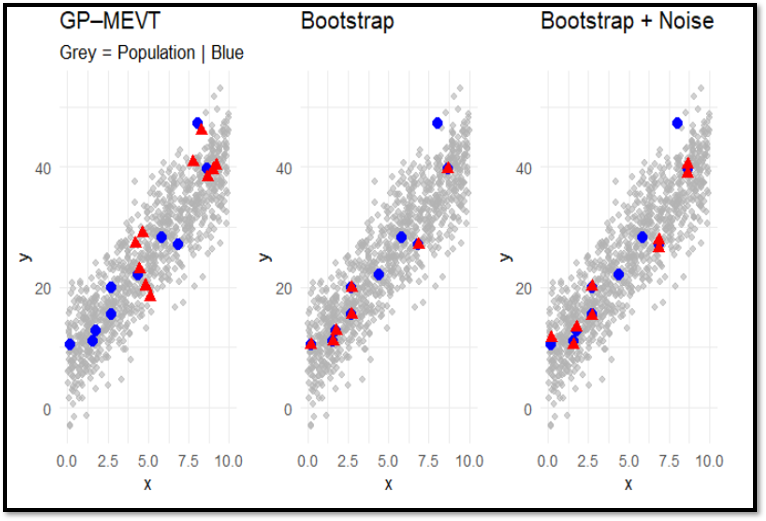} &
\includegraphics[width=0.30\textwidth]{Legend.png} \\
\end{tabular}
\caption{Augmented Data Distribution for a Random Sample of size 10.}
\label{Fig4}
\end{center}
\end{figure}

Following the same evaluation framework established in the simulation study, we compared the effectiveness of the GP-MEVT approach against both the standard bootstrap method and the bootstrap method with added noise through a comprehensive simulation study of 1,000 runs. For each run, we augmented the initial sample of size 10 with 20 additional augmented observations, then fitted a simple LR model, and its performance was evaluated. As in the previous application, we restricted our analysis to scenarios where the original small sample violated at least one classical regression assumption, ensuring that our evaluation focused on realistic scenarios where augmentation is most needed. Our comparative assessment examined multiple dimensions of model quality: (1) the overall rate at which augmented models satisfied all classical regression assumptions simultaneously, (2) individual satisfaction rates for homoscedasticity, normality of residuals, and independence, (3) the accuracy and precision of parameter estimates relative to known population values, (4) model explanatory power as measured by $R^2$, and (5) model accuracy quantified through RMSE. This evaluation provides a comprehensive picture of each method's ability to produce statistically valid and practically useful regression models from small samples drawn from real-world data. The simulation results for the samples of size 10 are summarized in Table \ref{tab:secondary_sim}.

\begin{table}[h!]
\centering
\caption{Simulation Results}
\vspace{0.5cm}
\small
\begin{tabular}{lcccccccc}
\toprule
Data & Assumption & Mean R & Mean & Intercept & Slope & \multicolumn{3}{c}{Individual assumptions} \\
augmented & satisfaction & squared & RMSE & (SE) & (SE) & (Homo) & (Norm) & (Indep) \\
methods& Rate & & & & & & & \\
\midrule
GP-MEVT & 0.6710 & 0.7863 & 5.5730 & 8.0253 & 3.3886 & 858 & 929 & 850 \\
& & & & (2.4937) & (0.3961) & & & \\
Bootstrap & 0.1730 & 0.8037 & 5.5923 & 8.0200 & 3.3943 & 592 & 409 & 855 \\
& & & & (2.1020) & (0.3751) & & & \\
Bootstrap & 0.2120 & 0.8037 & 5.5755 & 7.9977 & 3.3899 & 581 & 543 & 835 \\
+ Noise & & & & (2.0900) & (0.3749) & & & \\
\bottomrule
\end{tabular}
\label{tab:secondary_sim}
\end{table}

For the secondary data model evaluation, GP-MEVT demonstrated clear and substantial superiority over both Bootstrap methods across all performance metrics. The most pronounced advantage appeared in the overall assumption satisfaction rate, where GP-MEVT achieved nearly four times the success rate of standard Bootstrap and more than three times that of Bootstrap with Noise. This dramatic improvement indicates that GP-MEVT's approach to data augmentation generates synthetic samples that fundamentally respect the statistical requirements of classical linear regression far more effectively than resampling-based methods.

The homoscedasticity assumption is critical for valid inference in LR. GP-MEVT demonstrates superior performance with 858 out of 1000 simulation runs (85.8\%) satisfying this assumption, compared to Bootstrap (592/1000, 59.2\%) and Bootstrap with Noise (581/1000, 58.1\%). This 26.6 percentage point improvement is particularly noteworthy because heteroscedasticity can severely bias SE estimates and invalidate hypothesis tests, even when parameter estimates remain unbiased. Perhaps the most dramatic improvement is observed in residual normality, where GP-MEVT achieves a satisfaction rate of 92.9\% (929/1000 runs), compared to Bootstrap's 40.9\% (409/1000) and Bootstrap with Noise's 54.3\% (543/1000). This represents more than a doubling of the normality satisfaction rate relative to the best-performing bootstrap method. The normality assumption is essential for the validity of t-tests and F-tests used in hypothesis testing, as well as for constructing accurate confidence intervals for regression parameters. The standard bootstrap method's poor performance (40.9\%) likely reflects its tendency to replicate the distributional properties of the original small sample, which may itself exhibit non-normal residuals due to sampling variability. While Bootstrap with Noise shows improvement over standard bootstrap (54.3\%), the addition of random noise alone is insufficient to restore normality consistently. All three methods demonstrate comparable and satisfactory performance regarding residual independence, with GP-MEVT achieving 85.0\% (850/1000), Bootstrap achieving 85.5\% (855/1000), and Bootstrap with Noise achieving 83.5\% (835/1000). Independence of residuals is crucial for preventing autocorrelation, which can lead to inefficient parameter estimates and biased SEs. The consistently high satisfaction rates across all methods indicate that the augmentation procedures do not introduce artificial dependencies into the augmented data.

The parameter estimation results reveal important insights into the statistical properties of each method. The population parameters are $\beta_0 = 7.8739$ for the intercept and $\beta_1 = 3.42261$ for the slope, based on the full dataset of 1000 observations. GP-MEVT produces an intercept estimate of 8.025, representing a bias of +0.151 units (1.9\% relative bias) from the population value. Bootstrap yields 8.020, with a bias of +0.146 (1.9\%), while Bootstrap with Noise gives 7.998, with a bias of +0.124 (1.6\%). All three methods show comparable point estimates that slightly overestimate the true intercept. The augmentation process partially corrects this bias, with all methods moving closer to the true value. For the slope parameter, GP-MEVT estimates 3.389, representing a bias of -0.034 units (-1.0\% relative bias) from the population value of 3.423. Bootstrap produces 3.394, with a bias of -0.029 (-0.8\%), while Bootstrap with Noise yields 3.390, with a bias of -0.033 (-1.0\%). All three methods demonstrate excellent accuracy in slope estimation, with biases less than 1\% in relative terms. This performance is particularly impressive given that the original small sample estimated the slope as 3.987, substantially overestimating the true parameter by 16.5\%.

The mean $R^2$ values reveal an interesting pattern. Bootstrap and Bootstrap with Noise produce values nearly identical to the population $R^2$, while GP-MEVT yields a slightly lower value. A high $R^2$ value is meaningless and potentially misleading if the model violates fundamental assumptions. The bootstrap methods' $R^2$ values may be high, but only 17-21\% of their models meet the validity requirements for inference. GP-MEVT, with its 67.1\% assumption satisfaction rate, provides reliable $R^2$ values that can be trusted for interpretation. GP-MEVT achieves a mean RMSE marginally lower than Bootstrap with Noise and slightly better than Bootstrap. These differences are minimal in practical terms, indicating comparable model fitting accuracy across all methods. The near-equivalence in RMSE is notable because it demonstrates that GP-MEVT achieves its superior assumption satisfaction without sacrificing model performance. This addresses a common concern in statistical modeling that enforcing assumptions might compromise model fit, but ignoring them can compromise the validity of the entire model. The results show that GP-MEVT successfully navigates the bias-variance trade-off, producing models that are both statistically valid and robust.

\section{Conclusion}

This study introduced the GP-MEVT method as a novel data augmentation approach for fitting LR models to small samples. Through comprehensive evaluation across simulated datasets with varying variance structures and a secondary dataset, GP-MEVT demonstrated a substantial superiority over traditional bootstrap-based methods across multiple critical dimensions of model performance. The most notable advantage of GP-MEVT lies in its ability to produce augmented datasets that satisfy classical regression assumptions. Across all applications, GP-MEVT achieved assumption satisfaction rates ranging from 47\% to 68\%, representing a better improvement over standard bootstrap and bootstrap with noise methods. This improvement was consistent across individual assumptions, with particularly notable enhancements in residual normality and homoscedasticity. These high satisfaction rates directly translate to more reliable hypothesis testing, valid confidence intervals, and trustworthy statistical inference, critical requirements often unmet when working with small samples. The consistently high satisfaction rates by the GP-MEVT observed for all methods suggest that the augmentation procedures do not create artificial dependencies in the augmented data. In terms of parameter estimation accuracy, GP-MEVT consistently produced intercept and slope estimates closer to the true population values, with relative biases typically under 2\%. Across both applications, GP-MEVT's parameter estimates exhibited comparable accuracy with bootstrap alternatives, effectively correcting the substantial biases present in the original small samples. While GP-MEVT sometimes reported slightly larger SEs than bootstrap methods, this reflects more honest uncertainty quantification rather than reduced efficiency. Thus, GP-MEVT's uncertainty estimates are more trustworthy for statistical inference. Regarding model performance, GP-MEVT achieved competitive or superior results as measured by RMSE, demonstrating that enforcing assumption satisfaction does not compromise model fit. Simultaneously, GP-MEVT's mean $R^2$ values remained consistently closer to population $R^2$ values than bootstrap methods, avoiding the overfitting tendency inherent in resampling approaches. This balance between statistical validity and model fitting accuracy addresses a critical challenge in small sample regression, which is achieving models that are both theoretically sound and practically useful. 

The figures related to the random samples show that GP-MEVT generates augmented observations that not only conform to the sample range but also extend toward regions of the true population (extrapolation). This behavior is particularly valuable for small samples where certain predictor domain regions may be poorly represented. In contrast, standard Bootstrap merely replicates existing sample points, producing augmented observations clustered tightly around the original data, a limitation that becomes especially problematic when the sample poorly represents the population or violates regression assumptions.  Bootstrap with Noise partially remedies this by dispersing augmented points outward through added random noise. However, this spreading occurs without regard to the underlying data structure or population characteristics, making it largely undirected.

However, even with GP-MEVT, approximately one-third of augmented models fail to satisfy all classical assumptions, suggesting that some small samples may be too poorly structured for augmentation alone to enable valid inference. In such cases, researchers should consider complementary approaches such as Bayesian methods with informative priors or explicitly acknowledge inferential limitations. Additionally, GP-MEVT consumes higher computational costs than simple bootstrap resampling due to GP fitting and extreme value boundary computation.

Beyond these specific methodological findings, this research underscores the broader importance of generating augmented data that is statistically coherent and structurally aligned with the underlying data-generating process. In regression modeling with limited samples, poorly constructed augmented data can introduce artificial patterns, exacerbate assumption violations, and distort parameter estimates, ultimately undermining statistical inference validity. The GP-MEVT approach demonstrates that augmentation methods which explicitly preserve linear structure, respect distributional characteristics, and incorporate controlled variability can substantially enhance both model reliability and interpretability.

In conclusion, GP-MEVT represents a significant advancement in small sample regression methodology, offering practitioners a principled framework for conducting valid statistical inference when sample size limitations are unavoidable. By achieving substantial improvements in assumption satisfaction, parameter estimation accuracy, and model performance compared to bootstrap alternatives, GP-MEVT enables more robust modeling, improved generalization, and more credible inferential outcomes in small sample analytical settings. Future research directions include extending the method to multiple LR contexts, developing adaptive procedures for determining optimal augmentation sample sizes, and investigating performance under more complex assumption violation patterns such as non-linear relationships or correlated predictors.

\bibliographystyle{apalike}
\bibliography{References.bib}

@book{fox2015applied,
  title={Applied regression analysis and generalized linear models},
  author={Fox, John},
  year={2015},
  publisher={Sage publications}
}

@book{montgomery2012dirt,
  title={Dirt: The Erosion of Civilizations, with a New Preface},
  author={Montgomery, David R},
  year={2012},
  publisher={Univ of California press}
}

@book{johnson2011probability,
  title={Probability and statistics for computer science},
  author={Johnson, James L},
  year={2011},
  publisher={John Wiley \& Sons}
}

@book{harrell2001regression,
  title={Regression modeling strategies: with applications to linear models, logistic regression, and survival analysis},
  author={Harrell, Frank E and others},
  volume={608},
  year={2001},
  publisher={Springer}
}

@book{wooldridge2016introductory,
  title={Introductory Econometrics: A Modern Approach 6rd ed.},
  author={Wooldridge, Jeffrey M},
  year={2016},
  publisher={Cengage learning}
}

@article{tabachnick2013using,
  title={Using multivariate statistics (6. Bask{\i})},
  author={Tabachnick, Barbara G and Fidell, Linda S},
  journal={MA: Pearson},
  year={2013}
}

@book{cohen2013statistical,
  title={Statistical power analysis for the behavioral sciences},
  author={Cohen, Jacob},
  year={2013},
  publisher={routledge}
}

@article{button2013power,
  title={Power failure: why small sample size undermines the reliability of neuroscience},
  author={Button, Katherine S and Ioannidis, John PA and Mokrysz, Claire and Nosek, Brian A and Flint, Jonathan and Robinson, Emma SJ and Munaf{\`o}, Marcus R},
  journal={Nature reviews neuroscience},
  volume={14},
  number={5},
  pages={365--376},
  year={2013},
  publisher={Nature Publishing Group UK London}
}

@article{ lee2004evaluation,
  title={Evaluation of the Bayesian and maximum likelihood approaches in analyzing structural equation models with small sample sizes},
  author={Lee, Sik-Yum and Song, Xin-Yuan},
  journal={Multivariate behavioral research},
  volume={39},
  number={4},
  pages={653--686},
  year={2004},
  publisher={Taylor \& Francis}
}

@article{ peto1976design,
  title={Design and analysis of randomized clinical trials requiring prolonged observation of each patient. I. Introduction and design},
  author={Peto, Richard and Pike, MCetal and Armitage, Pet and Breslow, Norman E and Cox, David R and Howard, Simon V and Mantel, N and McPherson, K and Peto, J and Smith, PG},
  journal={British journal of cancer},
  volume={34},
  number={6},
  pages={585--612},
  year={1976},
  publisher={Nature Publishing Group}
}

@article{price2012small,
  title={Small sample properties of Bayesian multivariate autoregressive time series models},
  author={Price, Larry R},
  journal={Structural Equation Modeling: A Multidisciplinary Journal},
  volume={19},
  number={1},
  pages={51--64},
  year={2012},
  publisher={Taylor \& Francis}
}

@article{scheines1999bayesian,
  title={Bayesian estimation and testing of structural equation models},
  author={Scheines, Richard and Hoijtink, Herbert and Boomsma, Anne},
  journal={Psychometrika},
  volume={64},
  number={1},
  pages={37--52},
  year={1999},
  publisher={Springer-Verlag}
}

@book{woolson2002statistical,
  title={Statistical methods for the analysis of biomedical data},
  author={Woolson, Robert F and Clarke, William R},
  year={2002},
  publisher={John Wiley \& Sons}
}

@article{meuleman2015regression,
  title={Regression analysis: Assumptions and diagnostics},
  author={Meuleman, Bart and Loosveldt, Geert and Emonds, Viktor},
  journal={Regression analysis and causal inference},
  pages={83--110},
  year={2015},
  publisher={Sage Publications London}
}

@book{mcdonald2014handbook,
  title={Handbook of biological statistics},
  author={McDonald, John H},
  edition={3},
  year={2014},
  publisher={Sparky House Publishing, Baltimore, Maryland. USA}
}

@book{hastie2009elements,
  title={The elements of statistical learning: data mining, inference, and prediction},
  author={Hastie, Trevor and Tibshirani, Robert and Friedman, Jerome H and Friedman, Jerome H},
  volume={2},
  year={2009},
  publisher={Springer}
}

@book{james2013introduction,
  title={An introduction to statistical learning: with applications in R},
  author={James, Gareth and Witten, Daniela and Hastie, Trevor and Tibshirani, Robert and others},
  volume={103},
  year={2013},
  publisher={Springer}
}

@article{dietterich1995overfitting,
  title={Overfitting and undercomputing in machine learning},
  author={Dietterich, Tom},
  journal={ACM computing surveys (CSUR)},
  volume={27},
  number={3},
  pages={326--327},
  year={1995},
  publisher={ACM New York, NY, USA}
}

@book{efron1994introduction,
  title={An introduction to the bootstrap},
  author={Efron, Bradley and Tibshirani, Robert J},
  year={1994},
  publisher={Chapman and Hall/CRC}
}

@article{gelman2013bayesian,
  title={Bayesian data analysis third edition},
  author={Gelman, Andrew and Carlin, John B and Stern, Hal S and Dunson, David B and Vehtari, Aki and Rubin, Donald B},
  journal={Chapman and Hall/CRC},
  year={2013}
}

@article{goodfellow2016deep,
  title={Deep feedforward networks},
  author={Goodfellow, Ian and Bengio, Yoshua and Courville, Aaron},
  journal={Deep learning},
  volume={1},
  pages={161--217},
  year={2016},
  publisher={MIT press Cambridge, MA, USA}
}

@incollection{rasmussen2003gaussian,
  title={Gaussian processes in machine learning},
  author={Rasmussen, Carl Edward},
  booktitle={Summer school on machine learning},
  pages={63--71},
  year={2003},
  publisher={Springer}
}

@article{schulz2018tutorial,
  title={A tutorial on Gaussian process regression: Modelling, exploring, and exploiting functions},
  author={Schulz, Eric and Speekenbrink, Maarten and Krause, Andreas},
  journal={Journal of mathematical psychology},
  volume={85},
  pages={1--16},
  year={2018},
  publisher={Elsevier}
}

@article{wang2016knn,
  title={KNN-based Kalman filter: An efficient and non-stationary method for Gaussian process regression},
  author={Wang, Yali and Chaib-draa, Brahim},
  journal={Knowledge-Based Systems},
  volume={114},
  pages={148--155},
  year={2016},
  publisher={Elsevier}
}

@article{hox2012few,
  title={How few countries will do? Comparative survey analysis from a Bayesian perspective},
  author={Hox, Joop JCM and van de Schoot, Rens and Matthijsse, Suzette},
  journal={Survey Research Methods},
  volume={6},
  number={2},
  pages={87--93},
  year={2012}
}

@article{mcneish2016using,
  title={On using Bayesian methods to address small sample problems},
  author={McNeish, Daniel},
  journal={Structural Equation Modeling: A Multidisciplinary Journal},
  volume={23},
  number={5},
  pages={750--773},
  year={2016},
  publisher={Taylor \& Francis}
}

@article{sahinler2007bootstrap,
  title={Bootstrap and jackknife resampling algorithms for estimation of regression parameters},
  author={Sahinler, Suat and Topuz, Dervis},
  journal={Journal of Applied Quantitative Methods},
  volume={2},
  number={2},
  pages={188--199},
  year={2007}
}

@article{ma2018probabilistic,
  title={Probabilistic forecasting of landslide displacement accounting for epistemic uncertainty: a case study in the Three Gorges Reservoir area, China},
  author={Ma, Junwei and Tang, Huiming and Liu, Xiao and Wen, Tao and Zhang, Junrong and Tan, Qinwen and Fan, Zhiqiang},
  journal={Landslides},
  volume={15},
  pages={1145--1153},
  year={2018},
  publisher={Springer}
}

@article{wan2013hybrid,
  title={A hybrid approach for probabilistic forecasting of electricity price},
  author={Wan, Can and Xu, Zhao and Wang, Yelei and Dong, Zhao Yang and Wong, Kit Po},
  journal={IEEE Transactions on Smart Grid},
  volume={5},
  number={1},
  pages={463--470},
  year={2013},
  publisher={IEEE}
}

@article{nelson2014cluster,
  title={Cluster sampling: a pervasive, yet little recognized survey design in fisheries research},
  author={Nelson, Gary A},
  journal={Transactions of the American Fisheries Society},
  volume={143},
  number={4},
  pages={926--938},
  year={2014},
  publisher={Oxford University Press Oxford, UK}
}

@article{phaladiganon2011bootstrap,
  title={Bootstrap-based T 2 multivariate control charts},
  author={Phaladiganon, Poovich and Kim, Seoung Bum and Chen, Victoria CP and Baek, Jun-Geol and Park, Sun-Kyoung},
  journal={Communications in Statistics—Simulation and Computation{\textregistered}},
  volume={40},
  number={5},
  pages={645--662},
  year={2011},
  publisher={Taylor \& Francis}
}

@article{goldberg1997regression,
  title={Regression with input-dependent noise: A Gaussian process treatment},
  author={Goldberg, Paul and Williams, Christopher and Bishop, Christopher},
  journal={Advances in neural information processing systems},
  volume={10},
  year={1997}
}

@inproceedings{inproceedings,
author = {Lin, Yao-San and Tsai, Tung-I},
year = {2013},
month = {11},
pages = {463-468},
title = {Using Virtual Data Effects to Stabilize Pilot Run Neural Network Modeling},
volume = {26},
isbn = {978-1-4673-5247-5},
booktitle = {Proceedings of IEEE International Conference on Grey Systems and Intelligent Services, GSIS},
doi = {10.1109/GSIS.2013.6714828}
}

@article{schmidt2017predicting,
  title={Predicting the thermodynamic stability of solids combining density functional theory and machine learning},
  author={Schmidt, Jonathan and Shi, Jingming and Borlido, Pedro and Chen, Liming and Botti, Silvana and Marques, Miguel AL},
  journal={Chemistry of Materials},
  volume={29},
  number={12},
  pages={5090--5103},
  year={2017},
  publisher={ACS Publications}
}

@article{lee2016prediction,
  title={Prediction model of band gap for inorganic compounds by combination of density functional theory calculations and machine learning techniques},
  author={Lee, Joohwi and Seko, Atsuto and Shitara, Kazuki and Nakayama, Keita and Tanaka, Isao},
  journal={Physical Review B},
  volume={93},
  number={11},
  pages={115104},
  year={2016},
  publisher={APS}
}

@article{lowery1970comparison,
  title={A comparison of methods of fitting the double exponential distribution},
  author={Lowery, MD and Nash, JE},
  journal={Journal of Hydrology},
  volume={10},
  number={3},
  pages={259--275},
  year={1970},
  publisher={Elsevier}
}

\newpage
\section*{Appendix}

\renewcommand{\thefigure}{A\arabic{figure}}
\renewcommand{\thetable}{A\arabic{table}}

\setcounter{figure}{0}
\setcounter{table}{0}

\begin{table}[H]
\centering
\caption{Comparison of GP-MEVT, Bootstrap, and Bootstrap with Noise Methods for a Single Sample  for Scenario 1}
\scriptsize
\begin{tabular}{llrrr}
\toprule
Sample Size & Metric & GP-MEVT & Bootstrap & Bootstrap with Noise \\
\midrule
10 & R-squared & 0.5751 & 0.4632 & 0.4857 \\
& Assumption Satisfaction & TRUE & TRUE & TRUE \\
& Intercept & 2.4872 & 3.0612 & 2.9995 \\
& Slope & 5.8841 & 5.6657 & 5.5515 \\
\midrule
15 & R-squared & 0.3029 & 0.0457 & 0.0540 \\
& Assumption Satisfaction & TRUE & TRUE & TRUE \\
& Intercept & 2.2450 & 2.1536 & 2.2517 \\
& Slope & 4.0979 & 3.0716 & 3.2115 \\
\midrule
20 & R-squared & 0.0340 & 0.0494 & 0.0630 \\
& Assumption Satisfaction & TRUE & TRUE & TRUE \\
& Intercept & 5.3423 & 3.3864 & 3.4419 \\
& Slope & 10.9211 & 8.4627 & 8.6014 \\
\bottomrule
\end{tabular}
\label{tab:case1_single}
\end{table}

\begin{table}[h!]
\centering
\caption{Comparison of GP-MEVT, Bootstrap, and Bootstrap with Noise Methods for a Single Sample for Scenario 2}
\scriptsize
\begin{tabular}{llrrr}
\toprule
Sample Size & Metric & GP-MEVT & Bootstrap & Bootstrap with Noise \\
\midrule
10 & R-squared & 0.6121 & 0.5548 & 0.5718 \\
& Assumption Satisfaction & TRUE & FALSE & FALSE \\
& Intercept & 1.6436 & 2.1903 & 2.0922 \\
& Slope & 3.5822 & 4.1160 & 3.9315 \\
\midrule
15 & R-squared & 0.3065 & 0.4546 & 0.4316 \\
& Assumption Satisfaction & TRUE & FALSE & TRUE \\
& Intercept & 1.5789 & 1.5590 & 1.5329 \\
& Slope & 4.9710 & 2.7462 & 2.7002 \\
\midrule
20 & R-squared & 0.7182 & 0.4005 & 0.4202 \\
& Assumption Satisfaction & TRUE & TRUE & FALSE \\
& Intercept & 0.9656 & 1.1897 & 1.2603 \\
& Slope & 1.9856 & 2.9607 & 3.1362 \\
\bottomrule
\end{tabular}
\label{tab:case2_single}
\end{table}

\begin{table}[h!]
\centering
\caption{Comparison of GP-MEVT, Bootstrap, and Bootstrap with Noise Methods for a Single Sample for Scenario 3}
\scriptsize
\begin{tabular}{llrrr}
\toprule
Sample Size & Metric & GP-MEVT & Bootstrap & Bootstrap with Noise \\
\midrule
10 & R-squared & 0.8869 & 0.9455 & 0.9431 \\
& Assumption Satisfaction & TRUE & FALSE & FALSE \\
& Intercept & 0.3316 & 0.4817 & 0.4469 \\
& Slope & 1.0890 & 0.8501 & 0.7889 \\
\midrule
15 & R-squared & 0.7874 & 0.8626 & 0.8633 \\
& Assumption Satisfaction & TRUE & TRUE & TRUE \\
& Intercept & 0.5418 & 0.5963 & 0.6230 \\
& Slope & 1.1513 & 1.0609 & 1.1084 \\
\midrule
20 & R-squared & 0.7903 & 0.5527 & 0.5915 \\
& Assumption Satisfaction & TRUE & FALSE & FALSE \\
& Intercept & 1.0016 & 0.8608 & 0.9292 \\
& Slope & 1.7975 & 2.4324 & 2.6255 \\
\bottomrule
\end{tabular}
\label{tab:case3_single}
\end{table}

\end{document}